\documentclass[12pt,english,floatfix,showkeys,superscriptaddress,aps,prd,preprint]{revtex4}
\usepackage[latin1]{inputenc}
\usepackage[T1]{fontenc}
\usepackage{lmodern}
\usepackage{amsmath}
\usepackage{amssymb}
\usepackage{graphicx}
\usepackage{float}
\usepackage{esint}
\usepackage{longtable}
\usepackage{dcolumn}
\usepackage{babel}
\usepackage{csquotes}
\usepackage{color}
\usepackage{slashed}
\usepackage{simplewick}
\usepackage{tikz-feynman}
\usepackage[]{tikz-feynman}
\usepackage{amsmath,latexsym}

\usepackage{hyperref}
\hypersetup{
    colorlinks,
    citecolor=blue,
    filecolor=green,
    linkcolor=purple,
    urlcolor=red,
}

\usepackage{slashed}

\usepackage{hyperref}
\hypersetup{colorlinks,breaklinks,
			citecolor=[rgb]{0,0.0,1.0},
            urlcolor=[rgb]{0.0,0.0,1.0},
            linkcolor=[rgb]{0,0.5,0.9}}

\begin{document}

\title{Meson scattering in a Lorentz-violating scalar QED at finite temperature}

\author{M. C. Ara\'{u}jo}
\email{michelangelo@fisica.ufc.br}
\affiliation{Universidade Federal do Cear\'a (UFC), Departamento de F\'isica,\\ Campus do Pici, Fortaleza - CE, C.P. 6030, 60455-760 - Brazil.}
\author{R. V. Maluf}
\email{r.v.maluf@fisica.ufc.br}
\affiliation{Universidade Federal do Cear\'a (UFC), Departamento de F\'isica,\\ Campus do Pici, Fortaleza - CE, C.P. 6030, 60455-760 - Brazil.}



\date{\today}

\begin{abstract}

This paper investigates how the nonzero temperature affects the differential cross-section for mesons scattering in a Lorentz-violating extension of the scalar electrodynamics. We initially discuss some features of the model and extract the zero temperature Feynman rules. Temperature effects are introduced using the Thermo Field Dynamics (TFD) formalism. It is shown that the corrections induced on the meson scattering are very large in the high-temperature regime. Furthermore, our results also suggest that temperature effects may contribute to new constraints on the Lorentz-violating parameters.

\end{abstract}


\maketitle

\section{Introduction}\label{intro}

Despite the wide acceptance of the standard model (SM) of particle physics, it still leaves some open questions, such as the neutrino oscillation \cite{barger2012physics}, the hierarchy problem \cite{arkani1998hierarchy}, and the lack of a quantum description of gravity. As is known, the SM describes three of the four fundamental interactions (electromagnetic, weak and strong interactions), and it preserves Lorentz covariance and $CPT$ symmetry. Gravity is separately described by the General Relativity theory. However, Lorentz and $CPT$ violation could arise from an underlying theory combining gravity with quantum physics such as string theory \cite{Kostelecky:1988zi,Kostelecky:1991ak}, loop quantum gravity \cite{Gambini:1998it,Bojowald:2004bb}, Horava-Lifshitz gravity \cite{Horava:2009uw,Cognola:2016gjy}, or even in noncommutative spacetimes \cite{Carroll:2001ws}. Such theories lie on Planck scale, i.e. $\sim$ $10^{19}\mbox{GeV}$, seventeen magnitudes order greater than
the electroweak scale associated with the SM, making direct experimental verification difficult or even impossible nowadays. An alternative approach that aims to investigate signs of the Lorentz and $CPT$ symmetries violation, measurable at low energies, is described by the Standard Model Extension (SME) proposed by Colladay and Kosteleck\'{y} \cite{Colladay:1996iz,Colladay:1998fq}. The fundamental concept underlying the SME is to extend the Lagrangian density of SM to an effective field theory that includes all the possible Lorentz-violating (LV) terms that still preserve the gauge symmetries present in the SM \cite{tasson2014we,will7377living,hees2016tests}. Furthermore, any coefficient that violates $CPT$ also violates Lorentz symmetry, so the SME also provides a general framework for studying $CPT$ breaking \cite{Kostelecky:1988zi,Colladay:1996iz,Colladay:1998fq}. 

Later on, the SME was divided into two sectors: minimal and non-minimal sectors \cite{Kostelecky:2009zp,Kostelecky:2008bfz}. The first is the restriction of the full SME to operators of mass dimension $d \leq 4$ and it has been extensively studied in several contexts, such as radiative corrections \cite{jackiw1999radiatively,kostelecky2002one}, neutrino oscillation \cite{adamson2008testing,adamson2010search,Katori:2012pe,MiniBooNE:2011pix}, Euler-Heisenberg effective action \cite{Furtado:2014cja}, and gravitational context \cite{Accioly:2016yht,Kostelecky:2016kkn}. In its turn, the non-minimal sector 
incorporates all arbitrary mass dimension operators, and although it leads to non-renormalizable theories, it has been also played a central role in formal contexts such as string theory \cite{Kostelecky:1988zi,Kostelecky:1991ak}, Riemann-Finsler geometry \cite{Kostelecky:2011qz,Schreck:2019mmr,Reis:2017ay,Colladay:2017bon,Silva:2015ptj,Foster:2015yta,Russell:2015gwa,AlanKostelecky:2012yjr,Edwards:2018lsn}, classical-particle propagation \cite{Reis:2021ban}, supersymmetric models \cite{Lehum:2018tpi,Ferrari:2017rwk,Belich:2015qxa,Colladay:2010tx}, noncommutative quantum field theories \cite{Carroll:2001ws,Hayakawa:1999yt}, among others.

More recently, a general technique for treating $CPT$ violation in neutral-meson oscillations was presented by Edwards and Kosteleck\'{y} \cite{Edwards:2019lfb}. The primary idea was to treat a meson-antimeson system using a Lorentz-violating extension of the SM scalar sector in which is presented a general effective scalar field theory in any spacetime dimension. This model contains explicit perturbative spin-independent Lorentz violating operators of arbitrary mass dimension \cite{Edwards:2018lsn}. The spin-independence is crucial because the great majority of the fundamental particles of the SM have spin, being the Higgs boson the only example of a fundamental spinless particle in the SM.

In this paper, the differential cross section for meson scattering in the Lorentz violating extension of the scalar electrodynamics proposed in \cite{Edwards:2018lsn} is calculated at finite temperature. The Thermo Field Dynamics (TFD) formalism is used for the treatment at nonzero temperature. Such an analysis is important because it can help us to understand how temperature may contribute to a new class of constraints on Lorentz violation parameters, and by the fact that the search for Lorentz violation effects using astrophysics processes in very high temperature regime has been carried out in Ref. \cite{Friedman:2020bxa}.

This paper is organized as follows. In Sec. \ref{secdois}, the model itself and its respective Feynman rules at zero temperature are presented. In Sec. \ref{sectres}, the TFD formalism is introduced and the modifications on the Feynman rules due to temperature are discussed. In Sec. \ref{secquatro}, the differential cross section for mesons scattering in the Lorentz violating
extension of the scalar electrodynamics  is calculated at finite temperature. In Sec. \ref{conclusion} some final remarks are made.

\section{Lorentz-violating extension to scalar QED}\label{secdois}

In this section we present the model to the Lorentz-violating scalar QED recently proposed by Edwards and Kosteleck\'{y} \cite{Edwards:2018lsn}. The Lagrange density in four-dimensional flat spacetime which describes the interaction between mesons by exchanging photons is
\begin{eqnarray}\label{sqedlagrangiana}
\mathcal{L} &=& G^{\mu \nu}(D_{\mu}\phi)^{\ast}D_{\nu}\phi-m^2\phi^{\ast}\phi - \frac{i}{2}\left[ \hat{K}^{\mu}_a\phi^{\ast}D_{\mu}\phi - \hat{K}^{\mu}_a(D_{\mu}\phi)^{\ast}\phi \right] \nonumber\\
&-& \frac{1}{4}F_{\mu \nu}F^{\mu \nu} - \frac{\lambda}{4}(\phi^{\ast}\phi)^2,
\end{eqnarray} where $G^{\mu \nu} = \eta^{\mu \nu} + \hat{K}_c^{\mu \nu}$ is a tensor composed by Minkowski metric with signature $(1,-1,-1,-1)$ and a null trace constant tensor $\hat{K}^{\mu \nu}_c$. Also, $D_{\mu} = \partial_{\mu} - ieA_{\mu}$ is the usual covariant derivative allowing photon and complex scalar fields coupling, and $\hat{K}^{\mu}_a$ is another constant tensor. Lorentz violation effects are governed by the coefficients $\hat{K}^{\mu \nu}_c$ and $\hat{K}^{\mu}_a $ that are scalars under particle Lorentz transformations. Furthermore, they are also constants because they do not depend on coordinates. Therefore, energy-momentum is conserved. Moreover, the hermiticity of $\mathcal{L}$ requires that parameters can be taken as hermitian. 

As in standard scalar QED, we can directly read the usual Feynman rules (at zero temperature) directly from Eq. (\ref{sqedlagrangiana}), where
\begin{eqnarray}\label{feynmanrules1}
\begin{tikzpicture}[baseline=(a2.base)]
 \begin{feynman}
    \vertex (a1);
    \vertex [right=2.5cm of a1] (a2);
    \diagram* {
      (a1) -- [charged scalar, edge label=\(p\)] (a2),
      };
  \end{feynman}
 \end{tikzpicture} = \frac{i}{p^2-m^2}
\end{eqnarray}
\begin{eqnarray}\label{muqnudesenhoondprofoton}
 \begin{tikzpicture}[baseline=(b2.base)]
 \begin{feynman}
    \vertex (b1) {\(\mu\)};
    \vertex [right= 3cm of b1] (b2) {\(\nu\)};
    
    \diagram* {
      (b1) -- [photon,  edge label=\(q\)] (b2), 
    };
  \end{feynman}
 \end{tikzpicture} = - \frac{i\eta_{\mu \nu}}{q^2}
\end{eqnarray} are respectively the scalar and photon propagators, and
\begin{eqnarray}
\begin{tikzpicture}[baseline=(a2.base)]
 \begin{feynman}
    \vertex (a1);
    \vertex [dot, right=1.25 of a1] (a2)[label=90:\(p\)]{};
    \vertex [right=1.25 of a2] (a3);
    \diagram* {
      (a1) -- [scalar] (a2) -- [scalar] (a3),
      };
  \end{feynman}
 \end{tikzpicture} = i\hat{K}^{\mu \nu}_cp_{\mu}p_{\nu}
\end{eqnarray}
\begin{eqnarray}
\begin{tikzpicture}[baseline=(a2.base)]
 \begin{feynman}
    \vertex (a1);
    \vertex [right=2.5cm of a1] (a2);
    \diagram* {
      (a1) -- [scalar, insertion=0.5, edge label=\(p\)] (a2),
      };
  \end{feynman}
 \end{tikzpicture} = -i\hat{K}^{\mu}_ap_{\mu}
\end{eqnarray} are the insertions on scalar propagator due to Lorentz-violating parameters and 
\begin{eqnarray}
\begin{tikzpicture}[baseline=(b3.base)]
 \begin{feynman}
    \vertex (a1);
    \vertex[right=0.9 of a1] (a2);
    \vertex[right=0.9 of a2] (a3);
    \vertex[below=0.9 of a1] (b1);
    \vertex[right=0.9 of b1] (b2);
    \vertex[right=0.9 of b2] (b3);
    \vertex[below=0.9 of b1] (c1);
    \vertex[right=0.9 of c1] (c2);
    \vertex[right=0.9 of c2] (c3);
    \diagram* {
      (a1) -- [anti charged scalar] (b2) -- [anti charged scalar] (a3),
      (c1) -- [charged scalar] (b2) -- [charged scalar] (c3),
      };
  \end{feynman}
 \end{tikzpicture} = -i\lambda
\end{eqnarray}
\begin{eqnarray}
\begin{tikzpicture}[baseline=(b3.base)]
 \begin{feynman}
    \vertex (a1);
    \vertex[right=0.9 of a1] (a2);
    \vertex[right=0.9 of a2] (a3);
    \vertex[below=0.9 of a1] (b1);
    \vertex[right=0.9 of b1] (b2);
    \vertex[right=0.9 of b2] (b3);
    \vertex[below=0.9 of b1] (c1);
    \vertex[right=0.9 of c1] (c2);
    \vertex[right=0.9 of c2] (c3);
    \diagram* {
      (a1) -- [photon] (b2) -- [photon] (a3),
      (c1) -- [charged scalar] (b2) -- [charged scalar] (c3),
      };
  \end{feynman}
 \end{tikzpicture} = 2ie^2g^{\mu \nu}
\end{eqnarray}
\begin{eqnarray}
\begin{tikzpicture}[baseline=(b3.base)]
 \begin{feynman}
    \vertex (a1);
    \vertex[right=0.9 of a1] (a2);
    \vertex[right=0.9 of a2] (a3);
    \vertex[below=0.9 of a1] (b1);
    \vertex[right=0.9 of b1] (b2);
    \vertex[right=0.9 of b2] (b3);
    \vertex[below=0.9 of b1] (c1);
    \vertex[right=0.9 of c1] (c2);
    \vertex[right=0.9 of c2] (c3);
    \diagram* {
      (a1) -- [charged scalar, edge label=\(p\)] (b2),
      (b2) -- [photon] (b3),
      (c1) -- [anti charged scalar, edge label'=\(p'\)] (b2),
      };
  \end{feynman}
 \end{tikzpicture} = ieg^{\mu \nu}(p_{\mu} - p'_{\mu})
\end{eqnarray}
\begin{eqnarray}\label{diag11}
\begin{tikzpicture}[baseline=(b3.base)]
 \begin{feynman}
    \vertex (a1);
    \vertex[right=0.9 of a1] (a2);
    \vertex[right=0.9 of a2] (a3);
    \vertex[below=0.9 of a1] (b1);
    \vertex[dot, right=0.9 of b1] (b2){};
    \vertex[right=0.9 of b2] (b3);
    \vertex[below=0.9 of b1] (c1);
    \vertex[right=0.9 of c1] (c2);
    \vertex[right=0.9 of c2] (c3);
    \diagram* {
      (a1) -- [charged scalar, edge label=\(p\)] (b2),
      (b2) -- [photon] (b3),
      (c1) -- [anti charged scalar, edge label'=\(p'\)] (b2),
      };
  \end{feynman}
 \end{tikzpicture} = ie\hat{K}_c^{\mu \nu}(p_{\mu} - p'_{\mu})
\end{eqnarray}
\begin{eqnarray}\label{diag12}
\begin{tikzpicture}[baseline=(b3.base)]
 \begin{feynman}
    \vertex (a1);
    \vertex[right=0.9 of a1] (a2);
    \vertex[right=0.9 of a2] (a3);
    \vertex[below=0.9 of a1] (b1);
    \vertex[crossed dot, right=0.9 of b1] (b2){};
    \vertex[right=0.9 of b2] (b3);
    \vertex[below=0.9 of b1] (c1);
    \vertex[right=0.9 of c1] (c2);
    \vertex[right=0.9 of c2] (c3);
    \diagram* {
      (a1) -- [charged scalar, edge label=\(p\)] (b2),
      (b2) -- [photon] (b3),
      (c1) -- [anti charged scalar, edge label'=\(p'\)] (b2),
      };
  \end{feynman}
 \end{tikzpicture} = -ie\hat{K}^{\mu}_a
\end{eqnarray}
\begin{eqnarray}\label{diag13}
\begin{tikzpicture}[baseline=(b3.base)]
 \begin{feynman}
    \vertex (a1);
    \vertex[right=0.9 of a1] (a2);
    \vertex[right=0.9 of a2] (a3);
    \vertex[below=0.9 of a1] (b1);
    \vertex[dot, right=0.9 of b1] (b2){};
    \vertex[right=0.9 of b2] (b3);
    \vertex[below=0.9 of b1] (c1);
    \vertex[right=0.9 of c1] (c2);
    \vertex[right=0.9 of c2] (c3);
    \diagram* {
      (a1) -- [photon] (b2) -- [photon] (a3),
      (c1) -- [charged scalar] (b2) -- [charged scalar] (c3),
      };
  \end{feynman}
 \end{tikzpicture} = 2ie^2\hat{K}_c^{\mu \nu}
\end{eqnarray} are all the possible interaction vertices. As we can see in diagrams \eqref{diag11}, \eqref{diag12} and \eqref{diag13} new Lorentz-violating vertices are introduced by the parameters $\hat{K}^{\mu \nu}_c$ and $\hat{K}^{\mu}_a$.

\section{Thermofield dynamics formalism}\label{sectres}

This section is devoted to presenting the Thermo Field Dynamics (TFD) formalism and its main features \cite{Takahashi:1996zn,Umezawa:1982nv,Umezawa:1993yq,khanna2009thermal,Santana:1995np,Santana:1996np}. We started with a brief discussion of how Feynman's rules change. The TFD is a real-time formalism that allows the ensemble average of any operator $\mathcal{O}$ to be calculated as an expectation value in a thermal vacuum $|\, 0(\beta)\, \rangle$, i.e.
\begin{eqnarray}\label{valespobserovacuotermbeta}
\langle \, \mathcal{O}\, \rangle \equiv \langle \, 0(\beta) \, |\,  \mathcal{O}\, |\, 0(\beta) \rangle.
\end{eqnarray} We are going to consider $k_B = 1$ so that $\beta = T^{-1}$, with $T$ being the system temperature. An appropriate definition of $|\, 0(\beta)\, \rangle$  leads the TFD to be performed in a degrees of freedom doubled structure where for any operator $\mathcal{O}$ in the original physical space $\mathcal{S}$ is associated 
\begin{eqnarray}
\hat{\mathcal{O}} \equiv \mathcal{O} - \Tilde{\mathcal{O}},
\end{eqnarray}
 in the double-space $\hat{\mathcal{S}} = \mathcal{S}\otimes \Tilde{\mathcal{S}}$ where $\Tilde{\mathcal{O}}$ is an operator in the tilde-space $\Tilde{\mathcal{S}}$,  thought as a copy of $\mathcal{S}$. It is important to emphasize that all physical observables rest in the original (non-tilde) system. The following conjugation rules connect the tilde and non-tilde spaces \cite{khanna2009thermal}:
\begin{eqnarray}\label{rct1}
\widetilde{(\mathcal{O}_i\mathcal{O}_j)} &=& \tilde{\mathcal{O}_i}\tilde{\mathcal{O}_j} ,\\
\widetilde{(a\mathcal{O}_i+b\mathcal{O}_j)} &=& a^{\ast}\tilde{\mathcal{O}_i}+b^{\ast}\tilde{\mathcal{O}_j},\\
(\tilde{\mathcal{O}_i})^{\dagger} &=& (\mathcal{O}_i^{\dagger})^{\tilde{}},\\
\label{rct4}
\widetilde{(\tilde{\mathcal{O}_i)}} &=& \kappa \mathcal{O}_i,
\end{eqnarray} where $\kappa=1$ for bosons and $\kappa=-1$ for fermions. As a consequence of doubling it is possible to define thermal bosonic creation and annihilation operators that are related to tilde and non-tilde operators by Bogoliubov transformations
\begin{eqnarray}\label{oprtermoperzero1}
a(\vec{k},\beta) &=& u_B(\vec{k},\beta)a(\vec{k}) - v_B(\vec{k},\beta)\tilde{a}^{\dagger}(\vec{k}) ,\\ \label{oprtermoperzero2}
a^{\dagger}(\vec{k},\beta) &=& u_B(\vec{k},\beta)a^{\dagger}(\vec{k}) - v_B(\vec{k},\beta)\tilde{a}(\vec{k}), \\ \label{oprtermoperzero3}
\tilde{a}(\vec{k},\beta) &=& u_B(\vec{k},\beta)\tilde{a}(\vec{k}) - v_B(\vec{k},\beta)a^{\dagger}(\vec{k}),\\ \label{oprtermoperzero4}
\tilde{a}^{\dagger}(\vec{k},\beta) &=& u_B(\vec{k},\beta)\tilde{a}^{\dagger}(\vec{k}) - v_B(\vec{k},\beta)a(\vec{k}),
\end{eqnarray} where \begin{eqnarray}\label{funcaoubossoncosh1}
u_{B}(\vec{k},\beta) &=& \frac{e^{\beta E_{\vec{k}}/2}}{\sqrt{e^{\beta E_{\vec{k}}-1}}}, \\
\label{funcaoubossoncosh2}
v_{B}(\vec{k},\beta) &=& \frac{1}{\sqrt{e^{\beta E_{\vec{k}}-1}}},
\end{eqnarray} and 
\begin{eqnarray}
 \left[ a(\vec{k},\beta),a^{\dagger}(\vec{k'},\beta)\right] &=& (2\pi)^3\delta^3(\vec{k}-\vec{k'}),\\
 \left[ \tilde{a}(\vec{k},\beta),\tilde{a}^{\dagger}(\vec{k'},\beta)\right] &=& (2\pi)^3\delta^3(\vec{k}-\vec{k'}),
\end{eqnarray} with the other commutation relations being null.
Analogous expressions are possible for fermions \cite{khanna2009thermal}.

In the TFD formalism, the differential cross-section for two final-state particles in the center-of-mass frame is defined by  
\begin{eqnarray}\label{secdifbetacmsomaspinmhatcmbetaquadrado}
      \left( \frac{d \sigma }{d\Omega}   \right)_{\beta,cm} = \frac{1}{2E^2_{cm}}\frac{ | \vec{k} |}{16 \pi^2 E_{cm}}|\hat{\mathcal{M}}(\beta)|^2,
\end{eqnarray} where $\vec{k}$ is the 3-momentum of an outgoing particle, $E_{cm}$ is the center-of-mass energy, and 
\begin{eqnarray}\label{mhatbetammenosmtilbeta29}
      \hat{\mathcal{M}}(\beta) = \mathcal{M}(\beta) - \tilde{\mathcal{M}}(\beta),
\end{eqnarray} is the transition amplitude for the scattering, given by 
\begin{equation}\label{matrizshatbeta}
   \hat{\mathcal{M}}(\beta) \equiv \,  _{\beta}\langle \, f\, |\hat{S}|\, i\, \rangle_{\beta}.
\end{equation}
The $\hat{S}$-matrix is defined as 
\begin{eqnarray}\label{operevotempozero}
   \hat{S} = \sum_{n=0}^{\infty}\frac{(-i)^n}{n!}\int\, d^4z_1\cdots d^z_n \mathcal{T}\left[ \hat{\mathcal{H}}_I(z_1)\cdots \hat{\mathcal{H}}_I(z_n) \right],
\end{eqnarray} with $\mathcal{T}$ being the time-evolution operator. As we can see, the initial and final states are temperature dependent, and to lead with them we have to write the interaction Hamiltonian in terms of the thermal operators by using the Bogoliubov transformations in \eqref{oprtermoperzero1}-\eqref{oprtermoperzero4}. 

From the above description, it can be shown that the only modification in the Feynman rules  \eqref{feynmanrules1}-\eqref{diag13} lies on the propagators, which become temperature dependent. The photon propagator, for example, is defined by 
\begin{eqnarray}\label{propfotontermaldef1}
\langle\, 0(\beta)\, |\mathcal{T}\left[ A_{\mu}(x)A_{\nu}(y) \right]|\, 0(\beta)\, \rangle ,
\end{eqnarray} where 
\begin{eqnarray}\label{campodefotonszertotemp}
 A_{\mu}(x) = \int \frac{d^3k}{(2\pi)^3}\, \frac{1}{\sqrt{2E_{\vec{k}}}}\, \sum_{m=0}^{3}\left[ a^m(\vec{k})\epsilon^m_{\mu}(k)e^{-ik\cdot x} + a^{m \dagger}(\vec{k})\epsilon^{\ast m}_{\mu}(k)e^{ik\cdot x} \right],
 \end{eqnarray} is the usual field solution for photons with the symbol $\epsilon^m_{\mu}(k)$ standing for the polarization vector. Using \eqref{campodefotonszertotemp} and the Bogoliubov transformations \eqref{oprtermoperzero1}-\eqref{oprtermoperzero4} we can easily calculate \eqref{propfotontermaldef1} and write
 \begin{eqnarray}\label{thermalphotonprop}
 \begin{tikzpicture}[baseline=(b2.base)]
 \begin{feynman}
    \vertex (b1) {\(\mu\)};
    \vertex [right= 3cm of b1] (b2) {\(\nu\)};
    
    \diagram* {
      (b1) -- [photon,  edge label=\(q\)] (b2), 
    };
  \end{feynman}
 \end{tikzpicture} = \left\lbrace \begin{array}{l}
     \frac{-i\eta_{\mu \nu}}{q^2} - 2\pi\, \eta_{\mu \nu}v_B^2(\vec{q},\beta)\delta(q^2)  \\
     \\
   \frac{i\eta_{\mu \nu}}{q^2} - 2\pi\, \eta_{\mu \nu}v_B^2(\vec{q},\beta)\delta(q^2)
 \end{array} \right.
\end{eqnarray} The upper result is the photon propagator in the non-tilde space and the bottom result is the analogous photon propagator in the tilde space, namely,
\begin{eqnarray}\label{analogoprothermaltil}
\langle\, 0(\beta)\, |\mathcal{T}[ \tilde{A}_{\mu}(x)\tilde{A}_{\nu}(y) ]|\, 0(\beta)\, \rangle.
\end{eqnarray} Similarly, we can show that
\begin{eqnarray}\label{feynmanrulesscalarthermoprop1}
\begin{tikzpicture}[baseline=(a2.base)]
 \begin{feynman}
    \vertex (a1);
    \vertex [right=2.5cm of a1] (a2);
    \diagram* {
      (a1) -- [charged scalar, edge label=\(p\)] (a2),
      };
  \end{feynman}
 \end{tikzpicture} = \left\lbrace \begin{array}{l}
     \frac{i}{p^2 - m^2} - 2\pi\, v^2_B(\vec{p},\beta)\delta(p^2-m^2) \\
     \\
   \frac{-i}{p^2 - m^2} - 2\pi\, v^2_B(\vec{p},\beta)\delta(p^2-m^2) 
 \end{array} \right. 
\end{eqnarray} is the scalar propagator in non-tilde space (upper) and tilde space (bottom). All the other Feynman rules for the tilde space can be found by using tilde conjugation rules. 

\section{Differential cross section for mesons scattering in a Lorentz-violating scalar QED}\label{secquatro}

Now, we are able to calculate the differential cross section at finite temperature by using the TFD formalism for the relevant process  
\begin{eqnarray}
M_a(p) + \Bar{M}_a(p') \rightarrow M_b(k) + \Bar{M}_b(k'),
\end{eqnarray}
in the Lorentz-violating scalar QED model defined by (\ref{sqedlagrangiana}). Here, $M_a$ stands for an a-type meson different from a b-type meson and $\Bar{M}$ stands for the antiparticles.   The tree-level Feynman diagrams common to both tilde and non-tilde spaces are
\begin{eqnarray}
\begin{tikzpicture}[baseline=(b3.base)]
 \begin{feynman}
    \vertex (a1);
    \vertex [right=1.2 of a1] (a2);
    \vertex [right=1.2 of a2] (a3);
    \vertex [right=1.2 of a3] (a4);
    \vertex [below=1.2 of a1] (b1);
    \vertex [right=1.2 of b1] (b2);
    \vertex [right=1.2 of b2] (b3);
    \vertex [right=1.2 of b3] (b4);
    \vertex [below=1.2 of b1] (c1);
    \vertex [right=1.2 of c1] (c2);
    \vertex [right=1.2 of c2] (c3);
    \vertex [right=1.2 of c3] (c4);

    \diagram* {
      (a1) -- [anti charged scalar, edge label'=\(p'\)] (b2),
      (b2) -- [photon, edge label=\(q\)] (b3), 
      (b3) -- [anti charged scalar, edge label'=\(k'\)] (a4),
      (c1) -- [charged scalar, edge label=\(p\)] (b2),
      (b3) -- [charged scalar, edge label=\(k\)] (c4),
    };
  \end{feynman}
 \end{tikzpicture} \hspace{0.2cm}
 \begin{tikzpicture}[baseline=(b3.base)]
 \begin{feynman}
    \vertex (a1);
    \vertex [right=1.2 of a1] (a2);
    \vertex [right=1.2 of a2] (a3);
    \vertex [right=1.2 of a3] (a4);
    \vertex [below=1.2 of a1] (b1);
    \vertex [dot, right=1.2 of b1] (b2){};
    \vertex [right=1.2 of b2] (b3);
    \vertex [right=1.2 of b3] (b4);
    \vertex [below=1.2 of b1] (c1);
    \vertex [right=1.2 of c1] (c2);
    \vertex [right=1.2 of c2] (c3);
    \vertex [right=1.2 of c3] (c4);

    \diagram* {
      (a1) -- [anti charged scalar, edge label'=\(p'\)] (b2),
      (b2) -- [photon, edge label=\(q\)] (b3), 
      (b3) -- [anti charged scalar, edge label'=\(k'\)] (a4),
      (c1) -- [charged scalar, edge label=\(p\)] (b2),
      (b3) -- [charged scalar, edge label=\(k\)] (c4),
    };
  \end{feynman}
 \end{tikzpicture} \hspace{2.2cm} \nonumber\\
 \\
 \begin{tikzpicture}[baseline=(b3.base)]
 \begin{feynman}
    \vertex (a1);
    \vertex [right=1.2 of a1] (a2);
    \vertex [right=1.2 of a2] (a3);
    \vertex [right=1.2 of a3] (a4);
    \vertex [below=1.2 of a1] (b1);
    \vertex [right=1.2 of b1] (b2);
    \vertex [dot, right=1.2 of b2] (b3){};
    \vertex [right=1.2 of b3] (b4);
    \vertex [below=1.2 of b1] (c1);
    \vertex [right=1.2 of c1] (c2);
    \vertex [right=1.2 of c2] (c3);
    \vertex [right=1.2 of c3] (c4);

    \diagram* {
      (a1) -- [anti charged scalar, edge label'=\(p'\)] (b2),
      (b2) -- [photon, edge label=\(q\)] (b3), 
      (b3) -- [anti charged scalar, edge label'=\(k'\)] (a4),
      (c1) -- [charged scalar, edge label=\(p\)] (b2),
      (b3) -- [charged scalar, edge label=\(k\)] (c4),
    };
  \end{feynman}
 \end{tikzpicture} \hspace{0.2cm}
 \begin{tikzpicture}[baseline=(b3.base)]
 \begin{feynman}
    \vertex (a1);
    \vertex [right=1.2 of a1] (a2);
    \vertex [right=1.2 of a2] (a3);
    \vertex [right=1.2 of a3] (a4);
    \vertex [below=1.2 of a1] (b1);
   \vertex [crossed dot, right=1.2 of b1] (b2){};
    \vertex [right=1.2 of b2] (b3);
    \vertex [right=1.2 of b3] (b4);
    \vertex [below=1.2 of b1] (c1);
    \vertex [right=1.2 of c1] (c2);
    \vertex [right=1.2 of c2] (c3);
    \vertex [right=1.2 of c3] (c4);

    \diagram* {
      (a1) -- [anti charged scalar, edge label'=\(p'\)] (b2),
      (b2) -- [photon, edge label=\(q\)] (b3), 
      (b3) -- [anti charged scalar, edge label'=\(k'\)] (a4),
      (c1) -- [charged scalar, edge label=\(p\)] (b2),
      (b3) -- [charged scalar, edge label=\(k\)] (c4),
    };
  \end{feynman}
 \end{tikzpicture} \hspace{0.2cm}
 \begin{tikzpicture}[baseline=(b3.base)]
 \begin{feynman}
    \vertex (a1);
    \vertex [right=1.2 of a1] (a2);
    \vertex [right=1.2 of a2] (a3);
    \vertex [right=1.2 of a3] (a4);
    \vertex [below=1.2 of a1] (b1);
    \vertex [right=1.2 of b1] (b2);
    \vertex [crossed dot, right=1.2 of b2] (b3){};
    \vertex [right=1.2 of b3] (b4);
    \vertex [below=1.2 of b1] (c1);
    \vertex [right=1.2 of c1] (c2);
    \vertex [right=1.2 of c2] (c3);
    \vertex [right=1.2 of c3] (c4);

    \diagram* {
      (a1) -- [anti charged scalar, edge label'=\(p'\)] (b2),
      (b2) -- [photon, edge label=\(q\)] (b3), 
      (b3) -- [anti charged scalar, edge label'=\(k'\)] (a4),
      (c1) -- [charged scalar, edge label=\(p\)] (b2),
      (b3) -- [charged scalar, edge label=\(k\)] (c4),
    };
  \end{feynman}
 \end{tikzpicture}\nonumber
\end{eqnarray} where we are considering only the first-order contributions on the Lorentz-violating parameters since they are expected to lead to small corrections of the standard results. Using the Feynman rules discussed in the previous section we can write
\begin{eqnarray}
\hat{\mathcal{M}}(\beta) &=& \frac{e^2}{q^2}\, v(\vec{p},\beta)\, v(\vec{p'},\beta)\, v(\vec{k},\beta)\, v(\vec{k'},\beta) \left[ e^{\beta (E_{\vec{p}} + E_{\vec{p'}}+E_{\vec{k}}+E_{\vec{k'}})/2}+1 \right] \nonumber\\
&\times& \left[ (p^{\nu}-p'^{\nu})(k_{\nu}-k'_{\nu}) + 2\hat{K}^{\mu \nu}_c(p_{\mu}-p'_{\mu})(k_{\nu}-k'_{\nu}) - \hat{K}^{\mu}_a(k_{\mu}+p_{\mu}-p'_{\mu}-k'_{\mu}) \right] \nonumber\\
&\times& \left\lbrace 1 - 2\, \pi \, i\,  q^2 \,  v^2(\vec{q},\beta)\, \delta(q^2)\left[ \frac{e^{\beta (E_{\vec{p}} + E_{\vec{p'}}+E_{\vec{k}}+E_{\vec{k'}})/2}-1}{e^{\beta (E_{\vec{p}} + E_{\vec{p'}}+E_{\vec{k}}+E_{\vec{k'}})/2}+1} \right] \right\rbrace,
\end{eqnarray} and the expression for $|\hat{\mathcal{M}}(\beta)|^2$ can be easily calculated. In this way, working in the center-of-mass frame, we arrive at the following expression   
\begin{eqnarray}
|\hat{\mathcal{M}}(\beta)|^2 &=& e^4\,  \frac{(e^{\beta E_{cm}}+1)^2}{(e^{\beta E_{cm}/2}-1)^4}\nonumber \\
&\times& \left\lbrace \left[ 1-\frac{m_b^2}{E^2} \right]\cos^2\theta + 4\sqrt{1-\frac{m_b^2}{E^2}}\cos \theta \left[ \frac{\hat{K}_c^{3j}(k_j-k'_j)}{E_{cm}} + \frac{\hat{K}^{j}_a(p_j-k'_j)}{E^2_{cm}} \right] \right\rbrace \nonumber\\
&\times& \left\lbrace 1+\left[ \frac{2\pi\, E^2_{cm}\, \delta(E^2_{cm})}{e^{\beta E_{cm}}+1}\right]^2 \right\rbrace,
\end{eqnarray} where
\begin{eqnarray}
p=(E,E\hat{z}); &\hspace{1cm}& k=(E,\vec{k}); \nonumber\\
p'=(E,-E\hat{z}); &\hspace{1cm}&  k'=(E,-\vec{k'}); \nonumber
\end{eqnarray} with $\hat{z}\cdot \vec{k} = \sqrt{E^2-m_b^2}\cos\theta$ and $E_{cm}=2E$. Here, $m_b$ is the mass of the b-type meson. Plugging this expression into the formula (\ref{secdifbetacmsomaspinmhatcmbetaquadrado}), the temperature-dependent differential cross section for distinct mesons can be written as
\begin{eqnarray}
\left( \frac{d\sigma}{d\Omega} \right)_{\beta,cm} &=& \mathcal{B}(\beta)\,  \frac{\alpha^2}{4E_{cm}^2}\left[ 1-\frac{m_b^2}{E^2} \right]^{3/2}\left\lbrace \cos^2\theta - 4\cos \theta \left[ \left( \hat{k}^{32}_c + \frac{\hat{k}^{2}_a}{2E_{cm}}\right)\sin \theta \right. \right. \nonumber\\
\nonumber\\
&+& \left. \left. \left( \hat{k}^{33}_c + \frac{\hat{k}^{3}_a}{2E_{cm}} \right)\cos \theta + \frac{\hat{k}^3_a}{2E_{cm}}\left( 1-\frac{m_b^2}{E^2} \right)^{-1/2} \right] \right\rbrace, 
\end{eqnarray}
where $\alpha = e^2/4\pi$ is the fine-structure constant and 
\begin{eqnarray}\label{casablanca40}
\mathcal{B}(\beta) = \frac{(e^{\beta E_{cm}}+1)^2}{(e^{\beta E_{cm}/2}-1)^4}\left\lbrace 1+\left[ \frac{2\pi\, E^2_{cm}\, \delta(E^2_{cm})}{e^{\beta E_{cm}}+1}\right]^2 \right\rbrace,
\end{eqnarray} is the thermal correction factor. As we can see, there is in $\mathcal{B}(\beta)$ a product of delta functions with identical arguments, namely $\left[ \delta(E_{cm}^2) \right]^2$. Such a term is a consequence of the propagators in TFD and represents a kind of apparent "pathologies" that are actually called pinch singularity \cite{Landsman:1986uw}. However, as mentioned in Refs. \cite{Landsman:1986uw,van2001introduction}, all these problems are avoided by working with the regularized form of delta-functions
and their derivatives, namely
\begin{eqnarray}\label{propriedadedaderivadadadeltaprop}
 2\pi i \frac{1}{n!}\frac{\partial^n}{\partial x^n }\delta (x) = \left( -\frac{1}{x+i\epsilon} \right)^{n+1} - \left( -\frac{1}{x-i\epsilon} \right)^{n+1}.
\end{eqnarray} If $\epsilon$ is kept finite, it can be shown  that potentially dangerous terms like that will cancel after all relevant diagrams have been taken into account.

Finally, integrating over $d\Omega$, we obtain the total cross section:
\begin{equation}
    \sigma_{\mbox{total}}=\mathcal{B}(\beta)\left(1-4\hat{k}^{33}_c-\frac{2\hat{k}^{3}_a}{E_{cm}} \right)\frac{\pi \alpha^{2}}{3 E_{cm}^{2}}\left[1-\frac{4 m_{b}^{2}}{E_{cm}^2}\right]^{3/2}.
\end{equation}

\begin{figure}[!ht]
    \centering
    \includegraphics[width=0.8\textwidth]{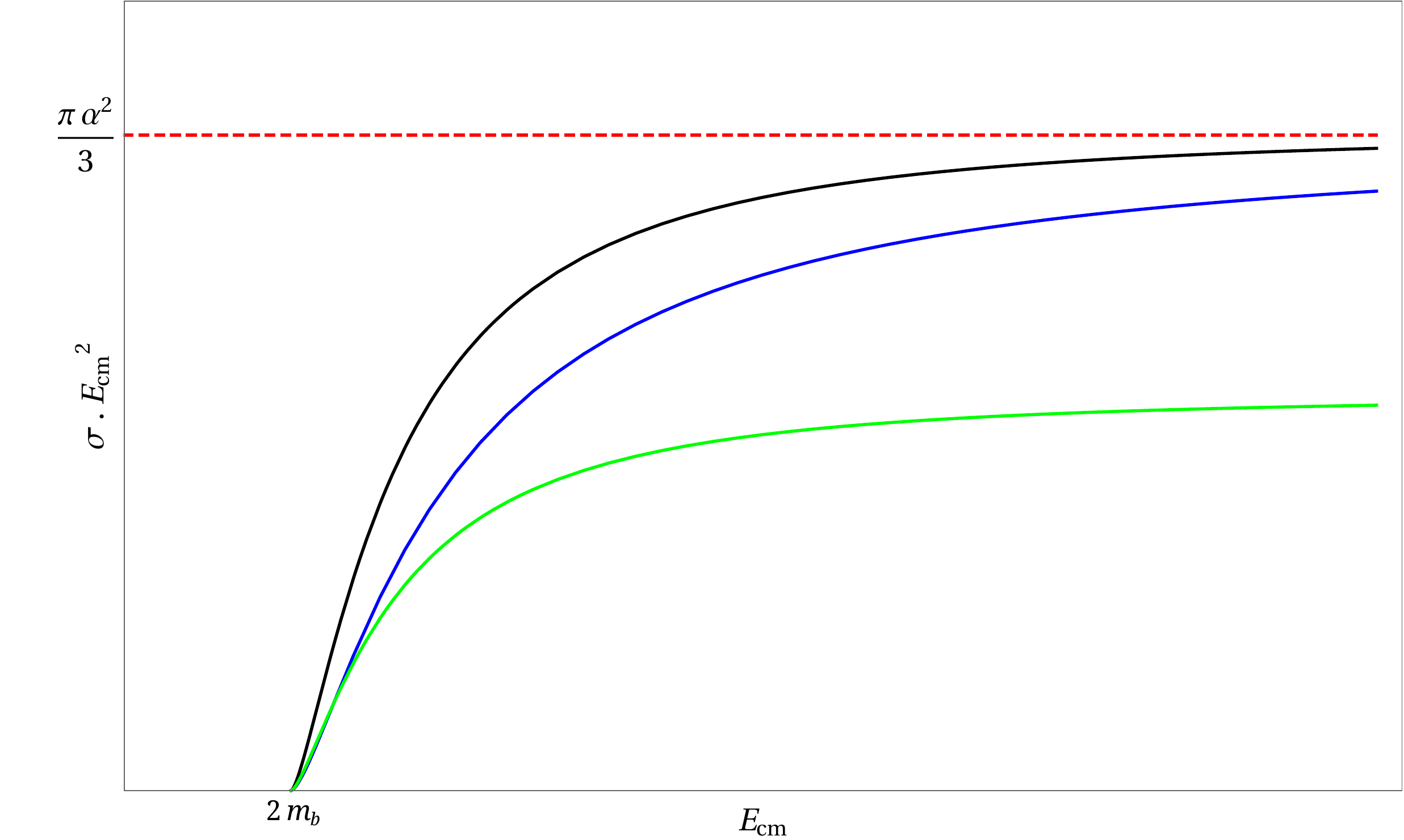} 
        \caption{Energy dependence of the total cross section at the zero temperature for the meson scattering to the Lorentz-violating scalar QED: Standard case (Black line), $\hat{k}^{33}_c=0$ and $\hat{k}^{3}_a=1/2$ (Blue line), $\hat{k}^{33}_c=1/10$ and $\hat{k}^{3}_a=0$ (Green line).}
    \label{fig1}
\end{figure}

Clearly, the behavior of the cross section with the temperature is dictated by the function $\mathcal{B}(\beta)$. In fact, this function goes to one in the zero temperature limit and the standard result for a Lorentz-violating scalar QED is recovered. In Fig. \ref{fig1}, we plot the total cross-section as a function of the energy of the center of mass at zero temperature. Note that the Lorentz symmetry violation, dictated by the coefficients $\hat{k}^{33}_c$ and $\hat{k}^{3}_a$, decreases the cross-section value, as can be seen from the blue and green curves. Furthermore, the thermal corrections to the cross section due to very high temperature turn to be extremely relevant once the function $\mathcal{B}(\beta)$ turns to be very large in this limit.
%

\section{Conclusion}\label{conclusion}

In this paper, the Lorentz violation effects on the meson scattering at finite temperature have been studied. The LV coefficients were introduced in the scalar QED through modifications only in the scalar sector of the model as proposed in Ref. \cite{Edwards:2018lsn}. Then, the first-order corrections due to the Lorentz symmetry violation to the scattering between distinct mesons were calculated. The finite temperature effects were introduced through the TFD formalism. As a result, the differential cross-section is modified by a temperature-dependent form factor $\mathcal{B}(\beta)$, and by the dependence of the Lorentz violation coefficients on the propagation direction of the mesons. The results indicate that in the high-temperature regime the thermal corrections become dominant, and in the zero-temperature limit we recover the results already obtained in the literature.


\section*{Acknowledgments}
\hspace{0.5cm} The authors thank Marco Schreck for valuable suggestions. They also thank the Funda\c{c}\~{a}o Cearense de Apoio ao Desenvolvimento Cient\'{i}fico e Tecnol\'{o}gico (FUNCAP), the Coordena\c{c}\~{a}o de Aperfei\c{c}oamento de Pessoal de N\'{i}vel Superior (CAPES), and the Conselho Nacional de Desenvolvimento Cient\'{i}fico e Tecnol\'{o}gico (CNPq), Grant no
311732/2021-6 (RVM) for financial support.


\end{document}